\documentstyle[aps,prl,epsf]{revtex}
\begin{document}
\title{Self force on static charges in Schwarzschild spacetime}
\author{Lior M. Burko\\
Theoretical Astrophysics \\
California Institute of Technology, Pasadena, California 91125}
\date{\today}

\maketitle

\begin{abstract}
We study the self forces acting on static scalar and electric test charges
in
the spacetime of a Schwarzschild black hole. The analysis is based on a
direct, local calculation of the self forces via mode decomposition, and
on two
independent regularization procedures: A spatially-extended particle
model method, and on a mode-sum regularization prescription.  
In all cases we find excellent agreement with the known exact results.  
\newline
\newline
PACS number(s): 04.25-g, 04.70.-s, 04.70.Bw
\end{abstract}

\section{Introduction and overview}
The problem of calculating the gravitational wave forms generated by
compact objects orbiting black holes is of crucial importance for
the detection
and the interpretation of observations by gravitational wave observatories
such as LISA \cite{lisa-93}. A major step towards the calculation of the
wave forms is the computation of the gravitational radiation reaction
forces, acting on the compact object. 
The generation of very
accurate templates for the waveforms detected from a system of a compact
object in orbit around a supermassive black hole is an extremely hard
task.
It is likely that one would need to have accurate templates for as
many as $10^5$ orbits. For such a system, accurate templates are necessary
for detection, because the predicted signal to noise ratio for LISA is
approximately of order $10$ for a $1$ year integration time.
Lack of accurate templates would results in a
loss of a factor of roughly the square root of the number of orbits
in sensitivity \cite{flanagan-hughes-98},
which would result in signal to noise ratio below the
detectability threshold.

Several methods have been suggested for the calculation of radiation
reaction.  One approach follows Dirac's method for obtaining the
Abraham-Lorentz-Dirac equation for an electric charge in arbitrary motion
in Minkowski spacetime \cite{dirac-38}. In that approach one imposes
local conservation laws on a tube surrounding the world line of the
particle, and integrates the conservation law across the tube, thus
obtaining the equations of motion, including the radiation reaction
effects. In Dirac's approach the infinities which are related to the
divergence of the particle's field on its world line are removed by a
simple mass renormalization. 
This method
was used by DeWitt and Brehme \cite{dewitt-brehme60} to generalize Dirac's
analysis for a general curved background. More recently, Mino, Sasaki, and
Tanaka used a similar method for the case of a massive particle coupled to
linearized gravity \cite{mino-97}. Quinn and Wald \cite{quinn-wald97} 
formulated recently an axiomatic approach for the calculation
of radiation reaction. In that approach, the infinities are removed by
comparing the forces in different spacetimes. 
However, at present it is unclear how to apply the
Quinn-Wald formal approach directly to practical calculations. The main
difficulty
arises from the calculation of the ``tail term'', which is difficult to
compute even in the slow motion, weak field limit
\cite{dewitt-dewitt64,quinn-wiseman}. 

Another approach is based on arguments relating to the balance of
quantities, which are constants of motion in the absence of radiation
reaction, specifically the energy and angular momentum. Such balance
arguments involve integration of the flux of an
otherwise conserved quantity over a boundary which consists 
of a distant sphere and the horizon of the black hole  
\cite{poisson-93,iyer-will93,iyer-will95,poisson-sasaki95,tagoshi-96,leonard-poisson97,gopakumar-97,quinn-wald99}.  
Although these methods are quite successful to
very high relativistic order in Schwarzschild spacetime, they are
problematic for the more interesting problem of motion in the spacetime of
a spinning black hole, because of an inherent difficulty: For the problem
of motion in the Schwarzschild spacetime the 
motion is completely determined by the rate of change of the energy and
the angular momentum, which are additive constants of motion in the
absence of radiation reaction. However, for general orbits in
the spacetime of a Kerr black hole there is a third constant of motion,
i.e., the Carter constant. The Carter constant is non-additive, and
consequently it cannot be obtained by methods which are based on
balance arguments. (Such methods can be
used for circular and equatorial orbits around a Kerr black hole, because 
then the evolution of the Carter constant is trivial: it is given
completely by the evolution of the energy and the azimuthal component of
the angular momentum.) In addition, such methods suffer also from other
difficulties \cite{wiseman}: they usually yield only the time average of
the radiation
reaction force, such that for any fast evolving system they would be
inherently inaccurate. In addition, they fail at obtaining the
conservative piece of the radiation reaction force.  

A different approach for the calculation of the gravitational radiation
reaction is based on a direct, local calculation of the self forces acting
on the
compact object. Obviously, knowledge of the instantaneous forces acting on
the orbiting object would allow for the calculation of the orbital 
evolution. Such a direct approach for the calculation of the self
force was suggested by Gal'tsov \cite{galtsov-82}. However, Gal'tsov's
approach is based on the radiative Green's function (i.e., the
``half-retarded minus half-advanced'' potential), which is not causal
in curved spacetime, because it requires the knowledge of the complete
future history of the object in motion \cite{ori-95}. A causal approach,
which is based on the retarded Green's function rather than on the
radiative one and consequently is more in the spirit of relativity theory, 
is much more desirable. 

Recently, a local approach for the calculation of the self force, which
is based on the retarded field, and on a Fourier-harmonic mode
decomposition of the field
and the self force, has been proposed \cite{ori-95,ori-97,paperI}. 
This approach has two very important
advantages: First, when the field is decomposed into modes, each mode
satisfies an ordinary differential equation rather than a partial one, and
consequently the solution for each mode is considerably simpler. Second,
and most importantly, each mode of the self force turns out to be finite.
Indeed, the total self force, which is obtained when one sums over all
modes, very frequently diverges, but this difficulty is met only at the
summation over all modes step: The treatment of the individual modes
is free from divergences. 

This approach was used by Ori
\cite{ori-95,ori-97} for the calculation of self forces acting on objects
in motion around a black hole. Ori suggested a regularization prescription
which
is based on the assumption that the divergent piece of the self force is
proportional to the four-acceleration of the charge. One can then use a
simple mass-renormalization procedure by re-defining the mass of the
particle to include the divergent piece. The self forces in general is 
expected to diverge. Therefore, a crucial ingredient for the calculation
of the self force is the regularization method which one uses. We note
that in any regularization prescription in the gravitational case (i.e.,
when the particle has a non-zero mass) one faces a gauge problem. This
difficulty, however, does not arise in the cases of scalar or
electromagnetic fields, because the force in these cases is gauge
independent. Therefore, consideration of scalar or electromagnetic charges
is of some value, as they correspond to easier cases, where much of the
difficulties related to self forces are already present, yet one does not
have to solve also the gauge problem. 

The self force is expected to be proportional to
the particle's (charge)$^2$. The correction to the orbit is therefore also
of order (charge)$^2$, and the correction to the self force is
consequently of order (charge)$^4$. When the charge of the particle is
much smaller than the mass of the black hole, this correction is
negligible. In this paper we shall study the self force only to leading
order, i.e., we shall study the self force to order (charge)$^2$.

We present in this paper two independent regularization
procedures for the self force, which are successful for the problem of
static charges (both scalar and electric) in the spacetime of a
Schwarzschild black hole.
(These procedures were also found to be successful for the regularization
of the radial component of the self force for scalar or electric charge
in uniform circular motion in flat spacetime
\cite{paperI,burko-unpublished}). 
We hope
that similar methods (or their generalizations) would be relevant also for
more complicated and realistic problems, e.g., the self forces acting on a
compact object in circular motion around a Schwarzschild black hole, and
ultimately, the self force on a compact
object in motion in a generic orbit around a Kerr black hole.

Let us consider first spatially-extended particles (we still assume that
the extension of the particles is smaller that the typical radius of
curvature and the typical scale of inhomogeneity of the field).  The
divergent piece of the self force, in addition to being
proportional to the four-acceleration, is also expected to be
inversely proportional to the spatial extension of the particle. In the
limit of a point-like particle, this is the source for the divergence
of the force. One should therefore be able to obtain a regularization
procedure by considering a spatially-extended model for the particle, and
then consider a sequence of smaller and smaller particles. The force
acting on the particles would increase like the inverse of their size, 
and by removing this piece of the force one can expect to obtain the
regularized self force, which is independent of the assumed internal
structure, in the limit of vanishing spatial extension. The
question of whether the regularized force depends of the way the 
point-like limit is taken is still an open question.  
A similar approach was 
used by Ori, who calculated the self forces acting on static
scalar and electric charges in Schwarzschild and on the axis of Kerr
black holes \cite{ori-unpublished}. In Schwarzschild, Ori used a 
dumbbell model, where the axis was aligned either radially
or tangentially. In Kerr, Ori used a radially-aligned dumbbell model. 
Whereas we use a mode-decomposition approach, which does not depend on the
availability of an exact solution, Ori used the exact solutions for the
scalar field or the electric potential, which are available for these
cases, in order to calculate the self forces.

We also consider a second, independent, regularization prescription. 
We consider a point-like particle. In that case the
sum over modes is expected in general to diverge. Ori has recently
suggested a mode-sum regularization prescription (MSRP) 
for the self force \cite{ori-private,barack-ori}. 
Although MSRP is not fully developed as yet, it has already
been shown to be valid for simple cases, such as a static scalar charge 
outside a Schwarzschild black hole. 
MSRP can possibly be generalized also for more complicated cases, such as 
massive particles in orbit around a Kerr black hole. If robust, MSRP  
can be of great importance for the generation of templates for the
detection of gravitational waves from compact objects in motion around
supermassive black holes.

The organization of this paper is the following. 
In Appendix A we describe very briefly the main ideas of MSRP, applied
for a scalar charge in Schwarzschild.   
In Section \ref{scalar} we 
discuss the self force acting on a static scalar charge in Schwarzschild
spacetime. The result has been obtained by independent methods: For a
minimally-coupled massless scalar field the self force vanishes
\cite{zelnikov-frolov82,wiseman,mayo}. 
It is our approach which is novel: Our calculation is based on a direct
computation of the self force mode by mode, followed by a 
summation over all modes, and
finally on two independent regularization procedures.  
One regularization procedure is based on a spatially-extended
particle model. We then consider the forces acting on a sequence of such
particles with decreasing spatial extensions, and remove the divergent
piece of the self force by a simple mass renormalization procedure. 
The other regularization procedure is based on MSRP. 
We find that both methods are successful in obtaining the correct
result.  
In Section \ref{electric} we consider the analogous problem of the 
self force acting on a static electric charge in Schwarzschild spacetime.
Also in this case, the result is not new. 
This problem has been considered by several authors: DeWitt and DeWitt 
\cite{dewitt-dewitt64} calculated the radiation damping forces (both
nonconservative and conservative) acting on a slowly moving electric
charge in the far-field regime, and found that there was a
repelling self force, which lowered the much stronger
gravitational pull of the black hole, and made a retrograde contribution
to the periastron precession.  
Vilenkin \cite{vilenkin-79} considered the
electric charge to be very far from the black hole (specifically, he
assumed the position of the charge to be at $r_0\gg M$, where $M$ is the
mass of the Schwarzschild black hole), and again found that there was a
repelling conservative self force. 
Smith and Will \cite{smith-will80} and
Frolov and Zel'nikov \cite{frolov-zelnikov80,zelnikov-frolov82} were able
to solve for the 
force exactly, for all positions of a static charge in Schwarzschild
spacetime, and found that the repulsive radial self force 
was $f^{\rm exact}_{\hat r}=e^2M/r^3$ (in the frame of a freely falling
observer who is instantaneously at rest at the position of the charge). 
Also in this case of a static electric charge we present a direct
approach for 
the calculation of the self force, which is based on
mode decomposition, summation over all modes, and force regularization
procedures similar to those we apply in the scalar case. 
In Section \ref{discussion} we summarize our methods and results.

\section{Static scalar charge}\label{scalar}
\subsection{Mode decomposition of the force}
Consider a point-like scalar test charge in the Schwarzschild spacetime,
held fixed by some external force. Our aim here is to calculate the
contribution of the self force to the total force needed for keeping it
fixed. The result is well known \cite{zelnikov-frolov82,wiseman,mayo}: 
the
contribution of the self force to the total force vanishes. The linearized
field equation of a minimally-coupled, massless scalar field $\Phi$ in the
Schwarzschild geometry, which is described by the line element 
$$\,ds^2=-\left(1-\frac{2M}{r}\right)\,dt^2+\left(1-\frac{2M}{r}\right)^{-1}
\,dr^2+r^2\,d\Omega^2,$$
where $\,d\Omega^2=\,d\theta^2+\sin^2\theta\,d\varphi^2 ,$
is given by
\begin{equation}
\nabla_{\mu}\nabla^{\mu}\Phi (x^{\alpha})=-4\pi\rho (x^{\alpha}),
\label{wave_scalar}
\end{equation}
where $\nabla_{\mu}$ denotes covariant differentiation, and where the
charge density 
\begin{equation}
\rho=q\int_{-\infty}^{\infty}\,d\tau\frac{\delta^{4}[x^{\mu}-
x_s^{\mu}(\tau)]}{\sqrt{-g}}.
\end{equation}
Here, $q$ is the charge, $\tau$ is its proper time, and $g$ is the metric
determinant. The mass of the black hole is denoted by $M$. 
The world line of the charge is given by $x_s^{\mu}(\tau)$. 
In what follows we use the usual Schwarzschild coordinates: The
radial Schwarzschild coordinate is defined such that spheres of radius $r$
have surface area $4\pi r^2$, and $t$ is the proper time of a static
observer at infinity. 
We take the charge to be on the
equatorial plane at the coordinates
$r=r_0$, $\varphi=0$, and $\theta=\pi/2$, without loss of generality.
(Because of the symmetry of the Schwarzschild geometry the coordinates
$\theta$ and $\varphi$ can be rotated such that these would be the
coordinates of any static charge at $r=r_0$.)   
Because of the staticity, the scalar field is independent of the time, and
we can decompose it into modes according to 
\begin{equation}
\Phi(r,\theta,\varphi)=\sum_{l=0}^{\infty}\sum_{m=-l}^{l}
\phi^{l}(r)Y^{lm}(\theta,\varphi),
\end{equation}
such that the LHS of the wave equation (\ref{wave_scalar}) is given by 
\begin{eqnarray}
\nabla_{\mu}\nabla^{\mu}\Phi(r,\theta,\varphi)&=&
\sum_{l=0}^{\infty}\sum_{m=-l}^{l}\left[\left(1-\frac{2M}{r}\right)\phi_{,rr}^l
\right. \nonumber \\
&+& \left. \frac{2}{r^2}(r-M)\phi_{,r}^l-\frac{l(l+1)}{r^2}\phi^l\right]
Y^{lm}.
\end{eqnarray}
The charge density is similarly decomposed into modes according to 
\begin{equation}
\rho=q\frac{\delta(r-r_0)}{r_0^2}\frac{1}{u^t(r_0)}
\sum_{l=0}^{\infty}\sum_{m=-l}^{l}Y^{lm\; *}(\frac{\pi}{2},0)
Y^{lm}(\theta,\varphi)
\end{equation}
where $u^{\alpha}$ is the four velocity of the charge, and a star denotes
complex conjugation. We thus find the radial equation for $\phi^{l}(r)$ to
be 
\begin{eqnarray}
\left(1-\frac{2M}{r}\right)\phi_{,rr}^l
+\frac{2}{r^2}(r-M)\phi_{,r}^l-\frac{l(l+1)}{r^2}\phi^l= \nonumber \\
-4\pi q\frac{\delta(r-r_0)}{r_0^2}\frac{1}{u^t(r_0)}
Y^{lm\; *}(\frac{\pi}{2},0).
\end{eqnarray}
To solve this equation we
transform to dimensionless harmonic coordinates, i.e., we define 
$\bar{r}\equiv (r-M)/M$. In the harmonic gauge the radial equation is
nothing but the Legendre equation\footnotemark 
\footnotetext{
That this should be the transformation is most easily seen from the 
following consideration. In the dimensionless coordinate $x=r/(2M)$ the
homogeneous equation is $(1-x)x\phi ''(x)+(1-2x)\phi '(x)
+l(l+1)\phi(x)=0$. This is a hypergeometric equation of the canonical form 
$x(1-x)\phi ''+[c-(a+b+1)x]\phi '-ab\phi=0$, for $a=-l$, $b=l+1$, and
$c=1$. As
$1-c=c-a-b$, we know from the theory of hypergeometric functions that the
homogeneous equation can be  transformed to Legendre's equation. The
variable of the hypergeometric equation $x$ is then related to the
variable of the Legendre equation by the transformation $x=(1+{\bar
r})/2$. In view of the
definition of $x$, we find that ${\bar  r}$ is nothing but the
dimensionless harmonic coordinate. We are then assured that
transformation to ${\bar r}$ would yield Legendre's equation with
solutions $P_l({\bar r})$ and $Q_l({\bar r})$ \cite{bateman}.}. We choose
the two independent solutions
of the corresponding homogeneous equation to be $P_l(\bar{r})$ and
$Q_l(\bar{r})$. The former is regular for $1<\bar{r}<\bar{r}_0$, and the
latter is regular for $\bar{r}>\bar{r}_0$. (Note that the horizon of the
black hole is located at $\bar{r}_{\rm horizon}=1$). The summation over
all
modes $m$ is readily done, and we thus write the
field at the point $(r,\theta,\varphi)$ due to a scalar charge $q$ at the
position $(r_s,\theta _s,\varphi _s)$ as 
\begin{eqnarray}
\Phi&=&
\frac{q}{M}\sqrt{1-\frac{2M}{r_s}}\sum_{l=0}^{\infty}(2l+1)
P_{l}(\cos\gamma)\nonumber \\
&\times &
\left[P_l\left(\frac{r_s-M}{M}\right)
Q_l\left(\frac{r-M}{M}\right)\Theta(r-r_s)\right. \nonumber \\
&+& \left. P_l\left(\frac{r-M}{M}\right)
Q_l\left(\frac{r_s-M}{M}\right)\Theta(r_s-r)\right].
\end{eqnarray}
Here, $\cos\gamma=\cos\theta\cos\theta_s+\sin\theta\sin\theta_s\cos
(\varphi-\varphi_s)$, and $\Theta(x)$ is the Heaviside step function,
i.e., $\Theta(x)=1$ for $x>0$ and $\Theta(x)=0$ for $x<0$. 
This solution for the scalar field $\Phi$ is regular both at the black
hole's event horizon and at infinity.   
In what follows as choose the angular coordinates
such that both the origin and the evaluation point of the field lie on the
equatorial plane, such that $\cos\gamma=\cos(\varphi-\varphi_s)$.

The force which a scalar field $\Psi$ exerts on a scalar charge $q'$ is
given by
$f_{\alpha}=q'\left(\Psi_{,\alpha}+u'_{\alpha}u'^{\beta}\Psi_{,\beta}\right)$, 
where the four-velocity $u'^{\alpha}$ is that of the charge $q'$. The
scalar field $\Psi$ can be any scalar field, in particular the self field
of the charge in question itself. 
In a sense, this is the scalar field analog of the electromagnetic Lorentz
force \cite{galtsov-82}. Because of the staticity of our problem the only 
component of $u'^{\alpha}$ which does not vanish is the temporal
component. However, the temporal derivative of the field vanishes, and
consequently the force is given only by $f_{\alpha}=q'\Psi_{,\alpha}$.
Because for scalar fields partial derivatives equal covariant derivatives,
this is, in fact, the covariant equation for the force.
Consider now two scalar charges, $q_1$ at $(r_1,\pi/2,\varphi_1)$ and 
$q_2$ at $(r_2,\pi/2,\varphi_2)$, and $r_2>r_1$. The force that $q_1$
exerts
on $q_2$ is given by 
\begin{eqnarray}
f_r^{12}(r_2)&=&\frac{q_1q_2}{M^2}\sqrt{1-\frac{2M}{r_1}}\sum_{l=0}^{\infty}
(2l+1)\nonumber \\
&\times &
P_l[\cos(\varphi_2-\varphi_1)]P_l\left(\frac{r_1-M}{M}\right)
Q_l'\left(\frac{r_2-M}{M}\right)
\end{eqnarray}
\begin{eqnarray}
f_{\varphi}^{12}(r_2)&=&
\frac{q_1q_2}{M}\sqrt{1-\frac{2M}{r_1}}\sum_{l=0}^{\infty}(2l+1)
\nonumber \\ 
&\times &
P_l\left(\frac{r_1-M}{M}\right)Q_l\left(\frac{r_2-M}{M}\right)
\frac{\,\partial P_l[\cos(\varphi_2-\varphi_1)]}
{\,\partial \varphi_2}.
\end{eqnarray}
Similarly, the force that $q_2$ exerts on $q_1$ is given by
\begin{eqnarray}
f_r^{21}(r_1)&=&\frac{q_1q_2}{M^2}\sqrt{1-\frac{2M}{r_2}}\sum_{l=0}^{\infty}
(2l+1)\nonumber \\ 
&\times &
P_l[\cos(\varphi_2-\varphi_1)]P_l'\left(\frac{r_1-M}{M}\right)
Q_l\left(\frac{r_2-M}{M}\right)
\end{eqnarray}
\begin{eqnarray}
f_{\varphi}^{21}(r_1)&=& 
\frac{q_1q_2}{M}\sqrt{1-\frac{2M}{r_2}}\sum_{l=0}^{\infty}(2l+1)
\nonumber \\
&\times &
P_l\left(\frac{r_1-M}{M}\right)Q_l\left(\frac{r_2-M}{M}\right)
\frac{\,\partial P_l[\cos(\varphi_2-\varphi_1)]} 
{\,\partial \varphi_1}.
\end{eqnarray}
Here, a prime denotes differentiation with respect to the argument. Each
of the force components is evaluated at the position of the charge on
which the force is exerted. We are interested to calculate the self force
acting on a point particle. Namely, we are interested in identifying $q_1$
with $q_2$. When this is done, one naturally finds that the total force
diverges (although each of the $l$ modes of the force is still finite).
Next, we describe two regularization procedures for the self force
acting on a point particle, which yield the desired result. 

\subsection{Regularization procedures}
\subsubsection{Extended particle model: Inclined dumbbell}
A well-known classical renormalization scheme is to consider a
spatially-extended particle model, and then consider the limit of
vanishing spatial extension, in the spirit of the classical
Abraham-Lorentz-Poincar\'{e} electron models. However, as is well known
\cite{quinn-wald97}, point-like particles are problematic in General
Relativity even to a greater extent than they are in electromagnetic
theory because of the non-linearity of the Einstein equations
\cite{geroch-traschen87}. Still,
in some sense, one can be hopeful that as the particle becomes smaller and
smaller, the deviation of its world line from a geodesic becomes
insensitive to the particle's internal structure.  
The simplest particle model
is a dumbbell model, consisting of two point-like charges at the two edges
of an uncharged rigid rod, whose length is smaller than the typical scales 
of the inhomogeneities of the gravitational and scalar or electric fields
(having in mind that we
shall later consider the limit of vanishing spatial extension). 
Although this is a very simplified model for a
particle, it can be simply generalized to more realistic models, bearing
in mind that a general extended (classical) object can be construed as
comprised of many point-like particles, and 
the self interaction of a general extended object can be
obtained by summing the contributions of all pairs of point-like
particles,
each pair being, in fact, a dumbbell. We shall thus model the particle to
be a dumbbell, with two equal charges $q_1=q_2=e/2$, $e$ being the total
charge of the particle. Because of the symmetry of the geometry, the
simplest configuration is to align the dumbbell axis in the radial
direction. That way, we still maintain axial symmetry, and the dumbbell
axis is aligned along a geodesic. However, we shall
see below that despite the fact that with a radial dumbbell axis one
indeed recovers the known and correct result for the self force, an 
important feature of a general extended particle model is missing,
specifically, the mass-renormalization aspect of the force regularization 
procedure. This happens because the coefficient of the divergent piece of
the bare force vanishes if the alignment of the dumbbell is radial (in the
scalar case). 
As we are interested primarily in the regularization procedure, we shall
consider here a more complicated case, where the dumbbell is not aligned
radially. Consequently, we shall take the
dumbbell axis to be inclined at some angle from the radial
direction\footnotemark \footnotetext{A
mathematical complication occurs if we take the dumbbell axis to be in the
$\partial/\,\partial\varphi$ direction. Specifically, in that case the
series expansion for the scalar field indeed converges, but not
absolutely. Consequently, one is not allowed to differentiate term by
term to obtain the force. This difficulty can most easily be illustrated 
in flat spacetime, where it already occurs: The electric scalar
potential due to a static unit point charge is given by 
$V=\sum_{l=0}^{\infty}\frac{r_<^l}{r_>^{l+1}}P_l(\cos\gamma)$. (See Ref. 
\cite{jackson} for details.) When the splitting is tangential, 
$r_<=r_>\equiv r$, and the potential is given simply by 
$V=(1/r)\sum_{l=0}^{\infty}P_l(\cos\gamma)$. It can be easily checked
that this series converges (although very slowly), but because of the
oscillations it does not converge absolutely. When the radial positions
of the source and the evaluation point are not equal there is an
additional attenuation, thanks to which the series converges absolutely.}. 
\begin{figure}
\input epsf
%
\centerline{ \epsfxsize 6.5cm
\epsfbox{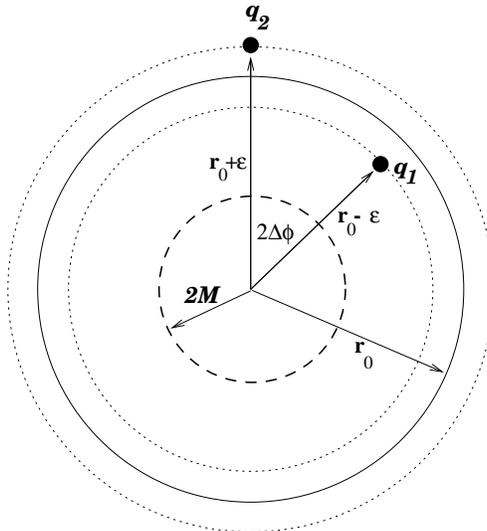}}
\caption{The geometry of the charge splitting in the equatorial plane: The
charge $e$ is
split into two half charges, $q_1=q_2=e/2$. The charge $q_1$ is placed at
$r_0-\epsilon$ and the charge $q_2$ is placed at $r_0+\epsilon$. The
total angular separation between $q_1$ and $q_2$ is $2\Delta\varphi$. The
horizon of the black hole is at $r=2M$.}
\label{fig_geometry} 
\end{figure}
Specifically, we take $r_2=r_0+\epsilon$, $r_1=r_0-\epsilon$, 
$\varphi_2=\Delta\varphi$, and $\varphi_1=-\Delta\varphi$. For
concreteness, we take $\Delta\varphi=\alpha\epsilon$, such that when we
make $\epsilon$ smaller, we also reduce $\Delta\varphi$ proportionally,
and we take $\epsilon \ll 2M$. 
Figure \ref{fig_geometry} illustrates the geometry of the charge splitting
in the equatorial plane (recall that the coordinates can always be rotated
such that the splitting is in the equatorial plane): The charges $q_1$ at
$r_0-\epsilon$ and $q_2$ at $r_0+\epsilon$ are also separated angularly by
an angle of $2\Delta\varphi$. When we take the limit $\epsilon\to 0$ we
simultaneously take the limit $\Delta\varphi\to 0$ too, such that the
point-like charge is located at the intersection of the circle of radius
$r_0$ and the bisector of the angle between $q_1$ and $q_2$.  

Let us consider only the radial force which acts on the
dumbbell. (Because the acceleration is purely radial, we expect only the
radial component of the self force to diverge.) 
The total (bare) self force which acts on the dumbbell is made of four
contributions. Schematically, 
\begin{equation}
f_r^{\rm total}=f_r^{12}+f_r^{21}+f_r^{11}+f_r^{22},
\end{equation} 
$f_{r}^{ij}$ being the radial component  of the force which the charge
$q_i$ exerts on the charge $q_j$. 
Let us consider this force in the point-like particle limit. 
Now, $f_r^{\rm total}$ is the self force on a {\it point-like} scalar
charge $e$. However, both $f_r^{11}$ and $f_r^{22}$ are equally well the
self forces on {\it point-like} scalar charges, which are identical to the
original charge $e$ in all respects, except for the fact that they each
have charge $e/2$. As
the self force is proportional to the charge squared, it implies that 
$f_r^{11}=f_r^{22}=f_r^{\rm total}/4$. Consequently, 
\begin{equation}
f_r^{\rm total}=2\left(f_r^{12}+f_r^{21}\right).
\label{total-force}
\end{equation}
Because we need to sum vector components, we have to perform the summation
at a common point, which for symmetry we choose to be
$(r_0,\pi/2,0)$\footnotemark \footnotetext{Note that the dumbbell is not
symmetric about this point, because the invariant distances from this 
point to the two edges are not equal. However, there is no particular
need for a symmetric dumbbell, and therefore we choose the model which
is the simplest mathematically.}. Specifically, we need to transport the
forces $f^{12}_{\mu}(r_2)$ and $f^{21}_{\mu}(r_1)$ to $(r_0,\pi/2,0)$
parallelly. 
For non-zero inclination angles the two edges of the dumbbell are not
separated by a geodesic of the background geometry. 
The final result for the self force
should of course be independent of the artificial spatial extension we
assume (i.e., independent of the internal structure of the particle), of 
the parallel-transport route, of the point where we sum the forces, 
and of the specific way at which we take
the point-like limit. It is still an open question whether the final
result depends on the way the
limit is taken. One might be worried about the introduction of ambiguities
due to the arbitrariness in the choice of the parallel-transport route.
Any ambiguity is of the order of the area enclosed by the two
routes we compare, times the curvature. The area is of order $\epsilon^2$,
and the curvature of order $M/r^3$, such that the ambiguity is of order
$\epsilon^2M/r^3$. Because the Coulomb components of the individual forces
cancel (see below), the leading order term in the total force is of order
$\epsilon^{-1}$, such that the ambiguity in the total force is of order
$\epsilon$, and vanishes in the limit $\epsilon\to 0$. 
That is, the final result is independent of the parallel-transport route. 
We note that the fact that the two edges of the dumbbell are not separated
by a geodesic is not a problem of principle, because for a general
extended object all pairs of the object's atoms interact, and most of them
are not separated by geodesics. 

We perform the parallel transport of $f^{12}_{\mu}(r_2)$ to $(r_0,\pi/2,0)$
in two steps: first, along the radial route
$(r_2,\pi/2,\varphi_2)\to (r_0,\pi/2,\varphi_2)$, and then along the 
tangential route 
$(r_0,\pi/2,\varphi_2)\to (r_0,\pi/2,0)$. Similarly, we parallel transport 
$f^{21}(r_1)$ from $(r_1,\pi/2,\varphi_1)$ first radially to 
$(r_0,\pi/2,\varphi_1)$ and then tangentially to $(r_0,\pi/2,0)$.
Note,
that
although we are eventually interested only in the radial component of the
self force, we need, in fact, to parallel transport both the radial and
the tangential components of the forces in the first sections of both
routes,
because when one parallel-transports a tangential component of a vector
tangentially, it acquires a radial component (already in flat space).
Another point to be made concerning the parallel transport is the
following: The individual forces can be expanded in a power series in
$\epsilon$,
where the leading term is proportional to $\epsilon^{-2}$. The self force
is of order unity, i.e., of order $\epsilon^{0}$. Therefore, one needs to
perform the parallel transport accurately at least to order
$\epsilon^{2}$. 
The parallel transports along
the radial routes is done as follows: The change in a covariant
component
of a vector in parallel transport along the
$\partial/\,\partial x^{\beta}$ direction satisfies 
$\delta V_{\alpha}=\Gamma^{\gamma}_{\alpha\beta}V_{\gamma}\,dx^{\beta}$,
where $\Gamma^{\gamma}_{\alpha\beta}$ are the connection coefficients,
which equal the Christoffel symbols of the second kind in General
Relativity. For the Schwarzschild geometry we find 
$\delta f_r=\Gamma^{r}_{rr}(r)f_r\,dr$ and 
$\delta f_{\varphi}=\Gamma^{\varphi}_{r\varphi}(r)f_{\varphi}\,dr$. 
Consequently, 
$\delta (\log f_r)=\,d\left(\log\sqrt{\frac{1}{1-2M/r}}\right)$ and
$\delta (\log f_{\varphi})=\,d(\log r)$, such that in the radial sections
of the parallel transports 
$f_r^{\rm new}=f_r^{\rm old}\sqrt{\frac{1-2M/r^{\rm old}}{1-2M/r^{\rm
new}}}$, and $f_{\varphi}^{\rm new}=f_{\varphi}^{\rm old}(r^{\rm 
new}/r^{\rm old})$. In the second sections of the parallel transportation,
the routes are tangential, such that 
$\delta f_r=\Gamma^{\varphi~}_{\varphi r}(r)f_{\varphi}\,d\varphi$. (We do
not
need to find the change in the tangential component of the force as we are
interested eventually only in the radial force.) That is, we need to
integrate $\delta f_r=(f_{\varphi}/r)\,d\varphi$. This can be done
straightforwardly, and we find 
\begin{eqnarray}
f_{r}^{\rm total}&=&\frac{1}{2}\frac{e^2}{M^2}\left\{
\sqrt{\frac{(1-2M/r_1)(1-2M/r_2)}{1-2M/r_0}}\sum_{l=0}^{\infty}(2l+1)
\right. 
\nonumber \\
&\times &\left.  
\left[P_l\left(\frac{r_1-M}{M}\right)Q_l'\left(\frac{r_2-M}{M}\right) 
\right. \right. \nonumber \\
&+& \left. \left. P_l'\left(\frac{r_1-M}{M}\right)
Q_l\left(\frac{r_2-M}{M}\right)\right]
P_{l}(\cos 2\Delta\varphi )\right.  \nonumber \\
&+&\left. 
M\left(\frac{1}{r_2}\sqrt{1-\frac{2M}{r_1}}+
\frac{1}{r_1}\sqrt{1-\frac{2M}{r_2}}\right)
\right. \nonumber \\
&\times & \left.
\sum_{l=0}^{\infty}(2l+1)
P_l\left(\frac{r_1-M}{M}\right)Q_l\left(\frac{r_2-M}{M}\right) \right. 
\nonumber \\
&\times & \left. 
\left[P_l(\cos\Delta\varphi)-P_l(\cos 2\Delta\varphi)\right] \right\}
\label{total-split}
\end{eqnarray}     
In Appendix B we describe briefly the numerical method we use for the
evaluation of the series. 
We evaluate this force for various values of $\epsilon$ (recall that
we take $\Delta\varphi$ to be proportional to $\epsilon$). We find that 
$f_r^{\rm total}$ diverges like $\epsilon^{-1}$, for very small values of 
$\epsilon$. This is indeed the expected behavior for the bare force.
Classical mass renormalization can be used for the regularization of the
bare force. Specifically, the divergent piece of the force is expected to
be proportional to the acceleration, such that it can be absorbed in the
mass of the particle. We use the exact solution for the scalar field
\cite{wiseman} 
\begin{eqnarray}
\Phi=q\sqrt{1-\frac{2M}{r_0}}
\left[(r-M)^2-2(r-M)(r_0-M)\cos\gamma+(r_0-M)^2-M^2\sin^2\gamma\right]
^{-1/2}, 
\end{eqnarray}
and sum the mutual forces of the two charges at the dumbbell's edges at a
common point, in the same way as above. Then, we expand the total force in
a power series in $\epsilon$, where the leading-order term is of order 
$\epsilon^{-1}$. This leading-order term is given by 
\begin{eqnarray}
f_r^{\rm div}&=&-e^2\frac{M}{r_0^2}\left(1-\frac{2M}{r_0}\right)^{-1}
\frac{1}{\epsilon}\left\{ \alpha^2
 \frac{r_0^2(r_0-M)}{M^3} \right. \nonumber \\ &\times& \left.  
\left(\alpha^2\frac{r_0^2}{M^2}+\frac{1}
{1-\frac{2M}{r_0}}\right)^{-3/2} 
-\frac{2}{r_0}\left(1-\frac{2M}{r_0}\right)^{3/2}
\right. \nonumber \\ 
&\times &  \left. 
\left[\frac{2}{\sqrt{4+\alpha^2\frac{r_0^2}{M^2}
\left(1-\frac{2M}{r_0}\right)}} \right. \right. \nonumber \\
&-& \left. \left. 
\frac{1}{\sqrt{1+\alpha^2\frac{r_0^2}{M^2}  
\left(1-\frac{2M}{r_0}\right)}}\right]\right\}.
\end{eqnarray}
Notice that the mass renormalization term depends on the value of the
parameter $\alpha$. That is, for different choices of $\alpha$ we re-scale
the mass by a different quantity. However, the renormalized, physical
self force is independent of $\alpha$, as should indeed be the case. 
We note that this mass renormalization procedure does not depend on the
availability of an exact solution. Below, in Section \ref{electric}, when
we discuss a similar mass renormalization for an electric charge, we do
not use the exact solution (although it is available). Instead, we use Eq. 
(\ref{griffiths}) for calculating the divergent piece of the self force. 
Even in cases where an equation analogous to Eq. (\ref{griffiths}) is not
available, the regularization procedure can still be done. 
In such a case one can extract the asymptotic divergence of the force at
small separation distances from the bare force, by finding the asymptotic
growth rate, and remove this piece from the bare force.  

\begin{figure}
\input epsf
\centerline{ \epsfxsize 8.5cm
\epsfbox{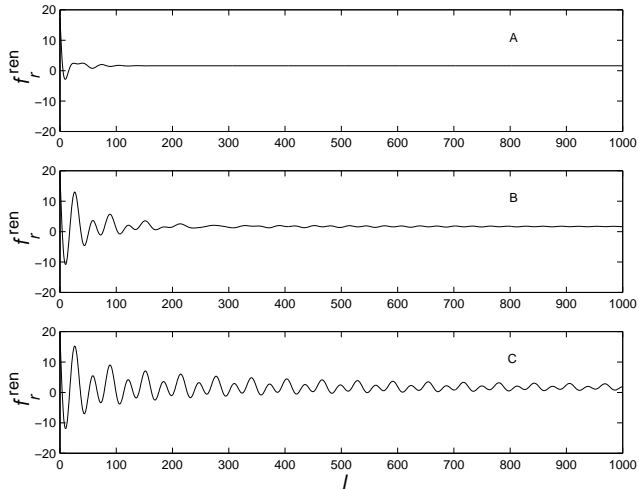}}
\caption{The behavior of the sum over modes of the renormalized force as a
function of $l$, for different values of the inclination parameter
$\alpha$. For all cases we take $r_0=2.1M$, and $\Delta\varphi=0.1$. 
Top panel (A): $\alpha=10M^{-1}$ (corresponding to $\epsilon=1\times
10^{-2}M$). 
Middle panel (B): $\alpha=10^{2}M^{-1}$ (corresponding to
$\epsilon=1\times
10^{-3}M$). 
Bottom panel (C): $\alpha=10^{3}M^{-1}$ (corresponding to
$\epsilon=1\times 
10^{-4}M$).
}
\label{fig0}
\end{figure}

We define the renormalized self force to be 
\begin{equation}
f^{\rm ren}_r\equiv f^{\rm total}_r-f^{\rm div}_r.
\end{equation}
The value of $f^{\rm ren}_r$ is of course a function of $\epsilon$, and we
need to take the self force in the limit of $\epsilon\to 0$. 
We find that the larger $\alpha$, the greater is the number of modes
over which we
need to sum until $f^{\rm ren}_r$ converges and the oscillations are
damped. Figure \ref{fig0} displays the behavior of the sums over modes up
to a certain value of the mode number $l$ as functions of $l$ for several
values of the inclination parameter $\alpha$. It is clear from Fig.
\ref{fig0} that for large inclination parameters one needs to sum over
many modes. In addition, we also find that, with fixed $\alpha$, the
number of modes one needs
to sum over scales like $\epsilon^{-1}$. When these two effects are
combined, one finds that it is very costly numerically to consider nearly
tangential splittings.
 
\begin{figure}
\input epsf
\centerline{ \epsfxsize 8.5cm
\epsfbox{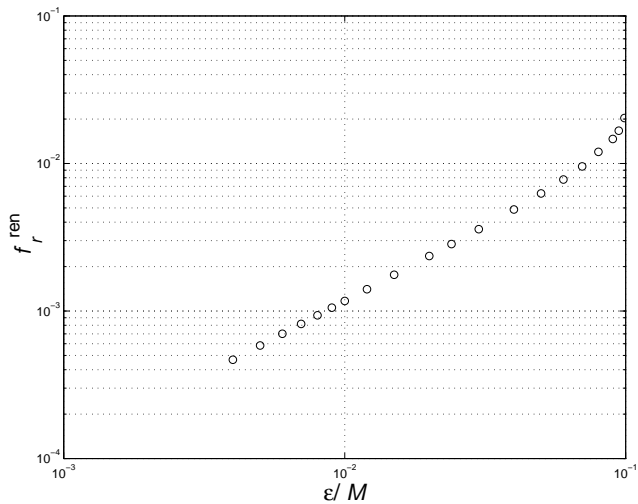}}
\caption{The renormalized self force $f_r^{\rm ren}$ as a function of the
spatial extension $\epsilon$. The charge is located at $r_0=2.1M$, and we
choose $\alpha=0.1$. We sum the $l$ modes here up to $l=2.8\times 10^{3}$.
}
\label{fig1} 
\end{figure}

Figure \ref{fig1} shows the renormalized force, i.e., 
$f^{\rm ren}_r\equiv f^{\rm total}_r-f^{\rm div}_r$ as a function of 
$\epsilon$ for a non-zero value of the inclination parameter $\alpha$. 
Similar results were obtained also for other values of $\alpha$ (but the
numer of modes we needed to sum over depended, of course, on the value of 
$\alpha$). The figure shows that for small spatial extension (small
values of $\epsilon$) the renormalized force is linear in $\epsilon$, such
that in the limit of vanishing spatial extension the force would equal
zero. Notice that we can see deviations from the linear law for large
spatial extensions. These deviations are expected, because the
renormalized force, when expanded in a power series in $\epsilon$,
contains contributions from all non-negative powers of $\epsilon$. The
self force is the force on a point-like particle, i.e., the force in the
limit $\epsilon\to 0$. Consequently, for any non-zero value of $\epsilon$
we have contributions also from all positive values of $\epsilon$, which
are dominated by the linear term in $\epsilon$ for small values of
$\epsilon$. 
In the special case where the alignment of the dumbbell axis is radial 
($\alpha=0$), we find that the divergent piece of the force
$f_{r}^{\rm div}$ vanishes
identically, such that $f_r^{\rm total}$ is already renormalized. In this
case we can sum the series in $f_r^{\rm total}$ analytically and find the
self force exactly. In fact, for any non-zero $\epsilon$ we find for a
radial dumbbell axis  
\begin{eqnarray}
f_r^{\rm total}&=&\frac{1}{2}\frac{e^2}{M^2}
\sqrt{\frac{\left(1-\frac{2M}{r_1}\right)\left(1-\frac{2M}{r_2}\right)}
{1-\frac{2M}{r_0}}}\sum_{l=0}^{\infty}(2l+1) 
\nonumber \\
&\times &
\left[
P_l\left(\frac{r_1-M}{M}\right)Q_l'\left(\frac{r_2-M}{M}\right)
\right. \nonumber \\ 
&+&
\left. 
P_l'\left(\frac{r_1-M}{M}\right)Q_l\left(\frac{r_2-M}{M}\right)\right]
\nonumber \\
&=&0,
\label{radial-split}
\end{eqnarray}
where we used the summation 
$\sum_{l=0}^{\infty}(2l+1)[P_l(x)Q_l'(y)+P_l'(x)Q_l(y)]=0$ for 
$y>x>1$. 

We thus find that if one considers a spatially-extended particle model for
the particle, one can obtain a finite result for the self force in the
limit of vanishing spatial extension, after performing a simple
mass-renormalization procedure, which agrees with the well-known exact
result 
\cite{zelnikov-frolov82,wiseman}. 

\subsubsection{Mode-sum regularization}
In this section we use MSRP in order to find the self force on a
point-like static scalar charge in Schwarzschild. MSRP is described
briefly in Appendix A, where the notation and definitions of the MSRP
parameters are given. We note that our
discussion here serves a dual purpose: First, it applies MSRP for a
specific case, and obtains non-trivial physical results. Seconds, because
our results can be compared with the final results for the self forces,
which are already known, it predicts values for the MSRP parameters, which
can then be tested analytically.  

\begin{figure}  
\input epsf
\centerline{ \epsfxsize 8.5cm
\epsfbox{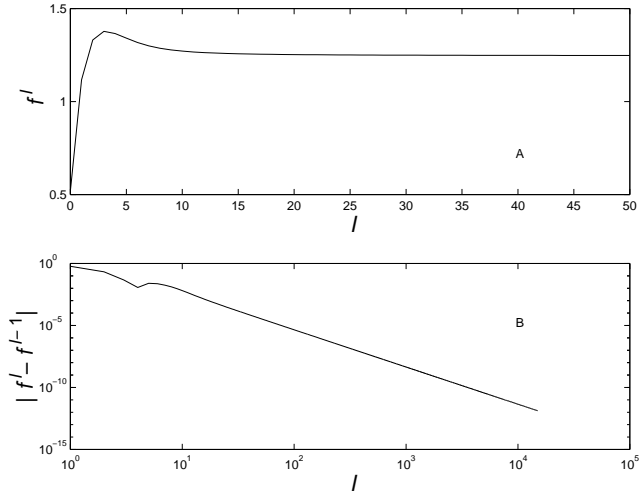}}
\caption{Behavior of the $l$ modes of the bare force for large values of
$l$. Top panel (A): $|{^{\rm bare}}f^l_r|$ as a function of $l$. Bottom
panel (B): 
$\left|{^{\rm bare}}f^l_r-{^{\rm bare}}f^{l-1}_r\right|$ as a function of
$l$. The scalar charge is
located at $r=2.1M$. 
}
\label{fig2}
\end{figure}

In the case of a point-like particle, we find from Eq. (\ref{total-split}) 
that the bare force is given by 
\begin{eqnarray}   
{^{\rm bare}}F_r&=&\frac{1}{2}\frac{e^2}{M^2}
\sqrt{1-\frac{2M}{r_0}}
\nonumber \\
&\times &
\sum_{l=0}^{\infty}(2l+1)\left[
P_l\left(\frac{r_0-M}{M}\right)Q_l'\left(\frac{r_0-M}{M}\right)
\right. \nonumber \\
&+& 
\left. 
P_l'\left(\frac{r_0-M}{M}\right)Q_l\left(\frac{r_0-M}{M}\right)\right].
\end{eqnarray}
Obviously, when this series is naively summed, the bare force 
diverges. In order to check the applicability of MSRP we first observe 
numerically that the $l$ modes of this forces ${^{\rm bare}}f_r^{l}$ 
approach a non-zero constant as $l\to\infty$, which we denote by
${^{\rm bare}}f_r^{\infty}$. 
Figure \ref{fig2} shows the convergence of the $l$ mode of
the bare force to a constant, as $l\to\infty$. The top panel of Fig.
\ref{fig2} shows ${^{\rm bare}}f^l_r$ as a function of $l$, for the first
few values of
the latter, and the bottom panel shows the difference between two
consecutive $l$ modes of the force as a function of $l$. We find that this
difference scales like $l^{-3}$ for large values of $l$, which implies
that indeed the series converges to a constant. 
This behavior implies that the MSRP parameters $a_r$ and $c_r$ vanish. 
(A non-zero value of $c_r$ implies that the difference between two
consecutive modes should
scale like $l^{-2}$.) 

\begin{figure}
\input epsf
\centerline{ \epsfxsize 8.5cm
\epsfbox{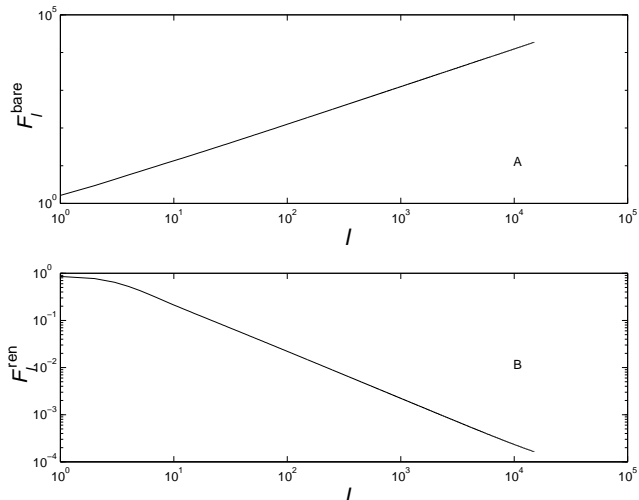}}
\caption{The bare force and the renormalized force as functions of the 
$l$ up to which we sum over the modes.
Top panel (A): ${^{\rm bare}}F_{r}^l(r_0)$ as a function of $l$. 
Bottom panel (B): ${^{\rm ren}}F_{r}^l(r_0)$ as a function of $l$.
For the renormalization procedure we use $l'=4.5\times 10^4$.
The scalar charge is located at $r_0=2.1M$.
}
\label{fig3}
\end{figure}

Because the $l$ modes of the bare force approach a non-zero constant 
as $l\to\infty$, it is clear that the sum over all modes diverges to
infinity. That is, the source for the divergence comes from the
contributions of the large-$l$ modes. 
Let us assume now that this divergence can be regularized by
removing the large-$l$ contributions. That is, we assume that the
large-$l$ contributions to the regularized self force die off with $l$.
The only sensible way to do that is to subtract the asymptotic value of
the modes (as $l\to\infty$) from all the modes of the bare force. Although 
this procedure yields a finite result for the self force, it is not {\it
a priori} clear whether that is the correct, physical result. However,
because the self force is already known, this can be checked, and
we can predict a value for a possible finite additional term for the
regularization procedure, which can be then be tested analytically using
MSRP. For obvious practical reasons, we do the summation over the modes 
only up to a finite
value of $l$. 
We denote the approximations of the bare and regularized forces (which are
obtained by summing over a finite number of modes) by 
${^{\rm bare}}F_{r}^l(r_0)$ and ${^{\rm ren}}F_{r}^l(r_0)$,
respectively.  
Then, we represent ${^{\rm bare}}f_r^{\infty}(r_0)$ by the $l'$ mode of
the force, for 
$l'$ much larger than the $l$ up to which we sum the series. In practice,
we find that $l'\approx 3l$ suffices to a very good accuracy. 
Specifically, 
\begin{equation}
{^{\rm tail}}F_{r}\approx{^{\rm
ren}}F_{r}^l(r_0)=\sum_{j=0}^{l}\left[{^{\rm
bare}}f_{r}^{j}(r_0)-{^{\rm bare}}f_{r}^{l'}(r_0)\right].
\end{equation}
Figure \ref{fig3} shows the bare force ${^{\rm bare}}F_{r}^l(r_0)$
and the renormalized force ${^{\rm ren}}F_{r}^l(r_0)$ as functions of $l$.
The bare force of course diverges for large values of $l$. However, 
figure \ref{fig3} implies that the renormalized force vanishes for large
$l$ like $l^{-1}$. Recall that the self force for this case is already
known to be zero \cite{zelnikov-frolov82,wiseman}. Consequently, we infer
that the value of the possible additional term for the regularization
procedure is zero. 
Indeed, MSRP yields for this particular case $d_r=0$, which agrees with
our result (see Appendix A). Because $d_r=0$, the regularized self force 
is given by MSRP to be simply
\begin{equation}
{^{\rm tail}}F_{r}=\sum_{l=0}^{\infty}\left[{^{\rm
bare}}f_{r}^{l}(r_0)-b_r(r_0)\right],
\end{equation}
where $b_r={^{\rm bare}}f_{r}^{l\to\infty}$, i.e., the regularization
procedure is reduced to subtracting the asymptotic value of the modes of
the bare force from all its modes, and then summation over all the modes. 

\begin{figure}
\input epsf
\centerline{ \epsfxsize 8.5cm
\epsfbox{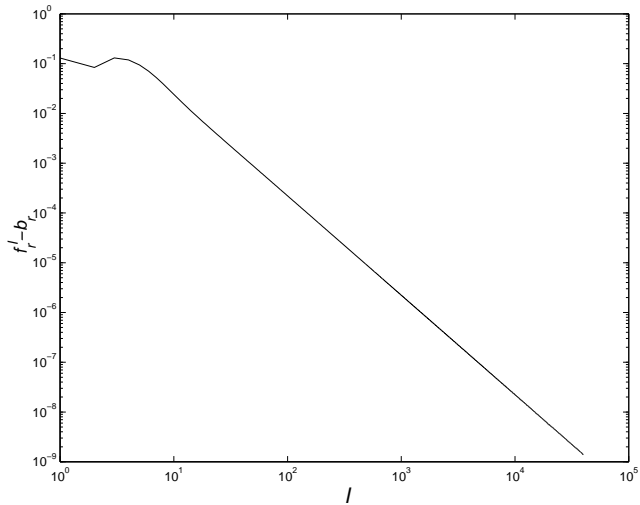}}
\caption{
The quantity ${^{\rm bare}}f^{l}_{r}(r_0)-b_r(r_0)$ as a function of
$l$. The scalar charge is located at $r_0=2.1M$.
}
\label{fig_msrp}  
\end{figure}

We can also check the prediction of MSRP for the exact value of $b_r$.
Recall that in this case $b_{r}=-[q^2/(2r^2)](1-M/r)/(1-2M/r)$, and that,
with $a_r=0$, MSRP predicts 
${^{\rm bare}}f^{l}_{r}\to b_r$ as $l\to\infty$.
Figure \ref{fig_msrp} displays the difference between ${^{\rm
bare}}f^{l}_{r}$ and $b_r$ as a function of $l$. This difference behaves
like $l^{-2}$ for large values of $l$. This asymptotic behavior  again
implies that $a_r=0$ and $c_r=0$, as we found above.  
For $r_0=2.1M$, we find this
difference to be $1.39\times 10^{-9}$ for $l=4\times 10^{4}$. This
agreement between the analytical prediction for $b_r$ and the value to
which the modes of the bare force approach at large values of the mode
number provides a strong support for the validity of MSRP.

\section{Static electric charge}\label{electric}
\subsection{Mode decomposition of the force}
An interesting case to study with our method is the case of a static
electric test charge in Schwarzschild spacetime. This is interesting
because it
is known that in this case the radial self force does not vanish. This can
give us two benefits. 
First, we can see whether our method can yield a correct
non-zero result (a zero result cannot reveal a wrong factor, say), and
second, we can use the exact expression for the result to evaluate the
error in our calculation. The exact result for the self force in this case
was found by Smith and Will \cite{smith-will80} and by Zel'nikov and
Frolov \cite{zelnikov-frolov82}. The field of a static electric charge in
the Schwarzschild spacetime was found in terms of a series expansion
solution by Cohen and Wald \cite{cohen-wald71} (see also
\cite{hanni-ruffini73,wiseman}.) 

The Maxwell equation in curved spacetime are given by
\begin{equation}
\nabla_{\nu}F^{\mu\nu}=4\pi j^{\mu}
\label{maxwell}
\end{equation}
where the Maxwell field strength tensor is given in terms of the
four-vector potential by 
$F_{\mu\nu}=A_{\nu ,\mu}-A_{\mu ,\nu}$, and where $j^{\mu}=\rho u^{\mu}$
is the four-current density, $\rho$ being the charge density. Because of
the staticity of
the problem (both the charge and the fixed background geometry are
static), all
spatial components of both the vector potential and the current density
vanish.  
The temporal component of Eq. (\ref{maxwell}) becomes
\begin{equation}
\frac{1}{\sqrt{-g}}\left(\sqrt{-g}g^{\nu\alpha}g^{tt}A_{t,\alpha}
\right)_{,\nu}=-4\pi j^t.
\end{equation}
In Schwarzschild coordinates this equation is explicitly written as 
\begin{eqnarray}
\left(r^2A_{t,r}\right)_{,r}&+& \left(1-\frac{2M}{r}\right)^{-1}\left[
\frac{1}{\sin\theta}\left(\sin\theta A_{t,\theta}\right)_{,\theta}
\right. \nonumber \\
&+& \left. \frac{1}{\sin^2\theta}A_{t,\varphi\varphi}\right]
=4\pi r^2j^t.
\end{eqnarray}
We next assume a series expansion of the form 
\begin{equation}
A_t(r,\theta,\varphi)=\sum_{l=0}^{\infty}\sum_{m=-l}^{l}
R^l(r)Y^{lm}(\theta,\varphi)
\end{equation}
and decompose the current density $j^{\mu}$ into modes 
\begin{equation}
j^t(r,\theta,\phi)=q\frac{\delta(r-r_0)}{r_0^2}
\sum_{l=0}^{\infty}\sum_{m=-l}^{l}Y^{lm\;*}(\frac{\pi}{2},0)
Y^{lm}(\theta,\varphi).
\end{equation}
This current density corresponds to a total charge $q$, as is evident from  
\begin{equation}
q=\int j^t(x^i)\sqrt{-g}\,d^3x^i.
\end{equation}
We thus find the radial equation to be
\begin{eqnarray}
\frac{d}{\,dr}\left[r^2\frac{\,dR_l(r)}{\,dr}\right] &-&
l(l+1)\left(1-\frac{2M}{r}\right)^{-1}R_l(r)
\nonumber \\ 
&=&
4\pi q\delta(r-r_0)Y^{lm\;*}(\frac{\pi}{2},0).
\label{inhomog-electric}
\end{eqnarray}
The basic functions which
solve the corresponding homogeneous
equation, with a convenient choice of normalization, are given by 
\cite{israel68,anderson-cohen70} 
\begin{equation}
R^{\infty}_{l}(r)=-\frac{(2l+1)!}{2^l(l+1)!l!M^{l+2}}(r-2M)Q'_{l}\left(
\frac{r-M}{M}\right) 
\end{equation}
\begin{equation}
R^{0}_{l}(r)=\frac{2^ll!(l-1)!M^{l-1}}{(2l)!}(r-2M)P'_{l}\left(
\frac{r-M}{M}\right)\;\;\;\;(l\ne 0),
\end{equation}
and $R^{0}_{0}(r)=1$. The Wronskian determinant of these two basic
solutions is \cite{cohen-wald71} $W_l(r)=-(2l+1)/r^2$. The solution of the
inhomogeneous equation (\ref{inhomog-electric}) is thus 
\begin{eqnarray}
R_l(r)&=&\frac{f(r_0)R_{l}^{\infty}(r_0)}{W_l(r_0)}R_{l}^{0}(r)\Theta(r_0-r)
\nonumber \\ 
&+&
\frac{f(r_0)R_{l}^{0}(r_0)}{W_l(r_0)}R_{l}^{\infty}(r)\Theta(r-r_0)
\end{eqnarray}
where 
\begin{equation}
f(r_0)=4\pi q\frac{1}{r_0^2}Y^{lm\;*}(\frac{\pi}{2},0) .
\end{equation}
The function  $R_l(r)$ is regular both at infinity and at the black hole's
event horizon. 
The summation over all modes $m$ is straightforward, and we find that the
$l$ mode $A_{t}^{l}$ satisfies 
\begin{eqnarray}
A_{t}^{l}&=&\frac{q}{M^3}\frac{2l+1}{l(l+1)}(r-2M)(r_0-2M)P_l(\cos\gamma)
\nonumber \\ 
&\times &
\left[P'_l\left(\frac{r-M}{M}\right)Q'_l\left(\frac{r_0-M}{M}\right)
\Theta(r_0-r) \right.
\nonumber \\
&+& \left. 
P'_l\left(\frac{r_0-M}{M}\right)Q'_l\left(\frac{r-M}{M}\right)
\Theta(r-r_0)\right]\;\;\;(l\ne 0).
\end{eqnarray}
For the monopole term ($l=0$) we find $A_{t}^{0}=-(q/r)\Theta(r-r_0)
-(q/r_0)\Theta(r_0-r)$.
Also in this case an exact solution is known \cite{wiseman}, which is 
\begin{eqnarray}
A_t=\frac{q}{r_0r}\left[M+
\frac{(r-M)(r_0-M)-M^2\cos\gamma}{\sqrt{(r-M)^2-2(r-M)(r_0-M)
\cos\gamma+(r_0-M)^2-M^2\sin^2\gamma}}\right].
\end{eqnarray}
The total covariant temporal component of the four-vector potential is
obtained by summing over all $l$ modes. The expression we thus find for 
$A_t$ is identical to the expression given in Ref. \cite{cohen-wald71} and
Ref. \cite{wiseman}. For the calculation of the force we need
only the gradient of $A_t$ with respect to $r$, which we simplify with the
differential equation which the Legendre functions satisfy. We find that 
\begin{eqnarray}
A_{t,r}&=&\frac{q}{r^2}\Theta(r-r_0)-\frac{q}{M^3}\frac{(r-2M)(r_0-2M)}{r}
\nonumber \\
&\times & \sum_{l=1}^{\infty}
\frac{2l+1}{l(l+1)}\left[
P'_l\left(\frac{r_0-M}{M}\right)Q'_l\left(\frac{r-M}{M}\right)\Theta(r-r_0)
\right. \nonumber \\
&+& \left. 
P'_l\left(\frac{r-M}{M}\right)Q'_l\left(\frac{r_0-M}{M}\right)\Theta(r_0-r)
\right] P_l(\cos\gamma) \nonumber \\
&+&
\frac{q}{M^2}\frac{r_0-2M}{r}\sum_{l=1}^{\infty}(2l+1)P_{l}(\cos\gamma) 
\nonumber \\ &\times &
\left[
P'_l\left(\frac{r_0-M}{M}\right)Q_l\left(\frac{r-M}{M}\right)\Theta(r-r_0)
\right. \nonumber \\
&+& \left.
P_l\left(\frac{r-M}{M}\right)Q'_l\left(\frac{r_0-M}{M}\right)\Theta(r_0-r)
\right].
\end{eqnarray}
We note that we did not include in this expression the terms proportional
to a delta function for the following reason. When the field is evaluated
at any point which is not the position of the charge, these terms are
zero.
When the evaluation point is at the position of the charge, the sum of the
terms proportional to a delta function vanishes. From this expression the
force is calculated according to the Lorentz force formula, specifically, 
$f^{\mu}=qF^{\mu}_{\;\;\nu}u^{\nu}$. Here, the only non-zero component of
the Maxwell field strength tensor is $F_{rt}=A_{t,r}$, and the only
non-zero component of the force is therefore the radial component.

\subsection{Regularization procedures}
\subsubsection{Extended particle model: radial dumbbell}
\begin{figure}
\input epsf 
\centerline{ \epsfxsize 8.5cm
\epsfbox{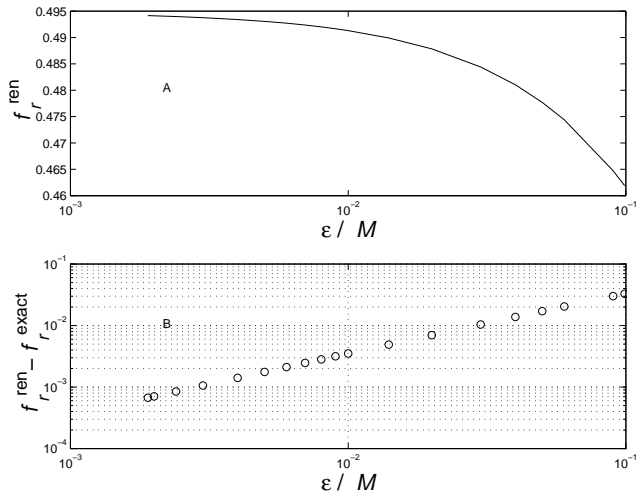}}
\caption{Top panel: The renormalized force as a function of
$\epsilon$. Bottom panel: $|f^{\rm ren}_r-f^{\rm exact}_r|$ as a
function
of $\epsilon/M$. The charge is located at $r=2.1M$. }
\label{fig4}
\end{figure}

Let us now assume for simplicity that the charges $q_1$ and $q_2$ are
separated only radially (there is no need for the more complicated
splitting we did in the scalar case, because in the electric case we still
need to perform mass
regularization even for radial splitting). 
As before, we sum the forces at a common point at $r_0$. After parallel
transporting the forces radially to the common point $r_0$, in the same
way it was
done above for the scalar case, we find that  
\begin{eqnarray}
{^{\rm bare}}F_r&=&\frac{e^2}{2}\frac{1}{\sqrt{1-\frac{2M}{r_0}}}
\left\{ \frac{1}{r_2^2} \right. \nonumber \\
&-& \left.
\sum_{l=1}^{\infty}(2l+1)\left[\left(\frac{1}{r_1}+\frac{1}{r_2}
\right)\frac{(r_1-2M)(r_2-2M)}{M^3l(l+1)}\right. \right. \nonumber \\ 
&\times & \left. \left. 
P'_l\left(\frac{r_1-M}{M}\right)
Q'_l\left(\frac{r_2-M}{M}\right) \right. \right. \nonumber \\
&-& \left. \left. \frac{r_1-2M}{r_2M^2}P'_l\left(\frac{r_1-M}{M}\right)
Q_l\left(\frac{r_2-M}{M}\right) \right. \right. \nonumber \\ 
&-& \left. \left. \frac{r_2-2M}{r_1M^2}P_l\left(\frac{r_1-M}{M}\right)
Q'_l\left(\frac{r_2-M}{M}\right)\right]\right\}
\label{force-electric}
\end{eqnarray}
We do not regularize this bare force in the limit $\epsilon\to 0$ with
the help of the exact solution because of the following. For an electric
dumbbell in arbitrary acceleration in flat spacetime, the divergent piece
of the self force is well known \cite{griffiths-owen83}, and is given by 
\begin{equation}
{\bf f}^{\rm div}=-\frac{e^2}{2d}\left[{\bf a}+\left({\bf a}\cdot{\hat
d}\right){\hat
d}\right].
\label{griffiths}
\end{equation}
Here, ${\bf a}$ is the three acceleration, ${\hat d}$ is a unit
three-vector in the direction of the dumbbell axis, and $d$ is the length
of the dumbbell axis. Note a factor of $2$ between this expression and Eq.
(57) of Ref. \cite{griffiths-owen83}, which is due to the fact that
according to Eq. (\ref{total-force}) the total force is {\it twice} the
sum of the two forces. One
would expect a similar expression to hold also in curved spacetime. 
In our problem, the dumbbell axis is aligned in the
radial direction, such that instead of ${\bf a}+\left({\bf a}\cdot{\hat d}
\right){\hat d}$ we would have $2a_r$. (The acceleration is only
radial.) We find 
$$a_r=\frac{M}{r_0^2}\left(1-\frac{2M}{r_0}\right)^{-1},$$ 
such that 
\begin{equation}
{^{\rm inst}}f_r=-\frac{e^2}{2}\frac{M}{r_0^2}\frac{1}
{\sqrt{1-\frac{2M}{r_0}}} \frac{1}{\epsilon}.
\end{equation}
Note, that the length of the dumbbell axis is given by the invariant
distance between $r_1$ and $r_2$. As in the scalar case, 
we perform mass renormalization by subtracting this
divergent piece of the force from the total force given by Eq. 
(\ref{force-electric}). Figure \ref{fig4} displays the renormalized force
as a function of $\epsilon$. We find that the renormalized force
approaches the {\it correct} finite value of
$f^{\rm exact}_r=e^2M/(r^3\sqrt{1-2M/r})$
\cite{smith-will80,zelnikov-frolov82} like $\epsilon$, as indeed we
expect. 

\subsubsection{Mode-sum regularization}
\begin{figure}
\input epsf
\centerline{ \epsfxsize 8.5cm
\epsfbox{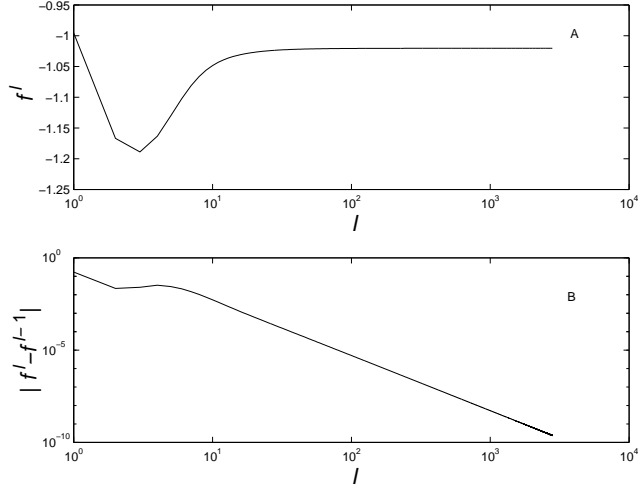}}
\caption{Behavior of the $l$ modes of the bare force for large values of
$l$. 
Top panel (A): The behavior of ${^{\rm bare}}F^{l}_{r}(r_0)$ as function
of $l$.
Bottom panel (B): 
$|{^{\rm bare}}F^{l}_{r}(r_0)-{^{\rm bare}}F^{l-1}_{r}(r_0)|$ 
as a function of $l$. The charge is located at $r=2.1M$. }
\label{fig5}
\end{figure}

As in the scalar case, we can also construe the charge as point-like, and
find from Eq. (\ref{force-electric}) the total bare radial self force to be 
given by  
\begin{eqnarray}
{^{\rm bare}}F_r(r_0)&=&\frac{e^2}{2}\left(1-\frac{2M}{r_0}\right)^{-1/2}
\left\{r_0^{-2}\right.\nonumber \\
&-&\left.2\frac{(r_0-2M)^2}{r_0M^3}\sum_{\ell=1}^{\infty}  
\frac{2\ell+1}{\ell(\ell+1)}P_{\ell}'\left(\frac{r_0-M}{M}\right)
Q_{\ell}'\left(\frac{r_0-M}{M}\right)\right. \nonumber \\
&+&\left. \frac{r_0-2M}{r_0M^2}\sum_{\ell=1}^{\infty}
(2\ell+1)\left[P_{\ell}'\left(\frac{r_0-M}{M}\right)
Q_{\ell}\left(\frac{r_0-M}{M}\right)\right.\right.\nonumber \\
&+& \left. \left.
P_{\ell}\left(\frac{r_0-M}{M}\right)
Q_{\ell}'\left(\frac{r_0-M}{M}\right)\right]\right\}.
\label{bare}
\end{eqnarray}
For calculation of the bare force, Eq. (\ref{bare}) can be re-written as 
\begin{eqnarray}
{^{\rm bare}}F_r(r_0)&=&\frac{e^2}{\sqrt{1-\frac{2M}{r_0}}}
\left[\frac{1}{r_0^2} - \frac{(r_0-2M)^2}{r_0M^3}
\right. \nonumber \\ &\times & \left. 
\sum_{l=1}^{\infty}\frac{2l+1}{l(l+1)}P'_l\left(\frac{r_0-M}{M}\right)
Q'_l\left(\frac{r_0-M}{M}\right)\right],
\label{bare1}
\end{eqnarray}
which simplifies the calculation. However, Eq. (\ref{bare1}) mixes the
contributions of the different modes. Although the regularization
procedure works also with this mixing, we shall consider below the
regularization procedure with the force as given by Eq. (\ref{bare}). 
We first check the behavior of the modes ${^{\rm bare}}f^{l}_{r}(r_0)$ as
$l\to\infty$. 
Figure \ref{fig5} shows that
indeed ${^{\rm bare}}f^{l}_{r}(r_0)$ approaches a constant, and that the
difference between two consecutive modes scales like $l^{-3}$ for large
values of $l$, in a similar way to the behavior of the modes for the
scalar case. Consequently, also for this case, we infer that $a_r=0$ and
that $c_r=0$. We emphasize that for the case of an electric charge these
parameters have not been calculated analytically, whereas in the scalar
case they have.  

In this case we also don't have prior knowledge about the value of the
parameter $d_r$. This is in general a serious problem, because without a
knowledge of $d_r$, the final result for  the self force is not
unambiguous. However, in this case we do have the final result from
independent approaches, such that we can, in fact, predict the value of
$d_r$. It remains work yet to be done to compute the values of 
$a_r$, $b_r$, $c_r$, and $d_r$ analytically for this case. 

As in the scalar case, we approximate the bare and the renormalized forces
by the sum over a finite number of modes, and denote them by 
${^{\rm bare}}F^{l}_{r}$ and ${^{\rm ren}}F^{l}_{r}$, respectively.  
We again define ${^{\rm ren}}F^{l}_{r}$ as in the scalar case, by
subtracting $f^{\infty}_r$ from each mode of the series. Figure 9  
shows the renormalized force ${^{\rm ren}}F^{l}_{r}$ and its difference
from $f^{\rm exact}_{r}$ as functions of the mode number $l$. We find that
${^{\rm ren}}F^{l}_{r}-f^{\rm  exact}_r$ approaches zero like $l^{-1}$ for
large values of $l$, such that we recover the results of Refs.
\cite{smith-will80,wiseman}, i.e., we find that the self force is a
repelling force, which is given by 
$f_r=q^2M/(r^3\sqrt{1-2M/r}).$ 
The asymptotic agreement of 
${^{\rm ren}}F^{l}_{r}$ and $f^{\rm  exact}_r$ imply that also for this
case $d_r=0$. This prediction can be tested analytically.

\begin{figure}
\input epsf
\centerline{ \epsfxsize 8.5cm
\epsfbox{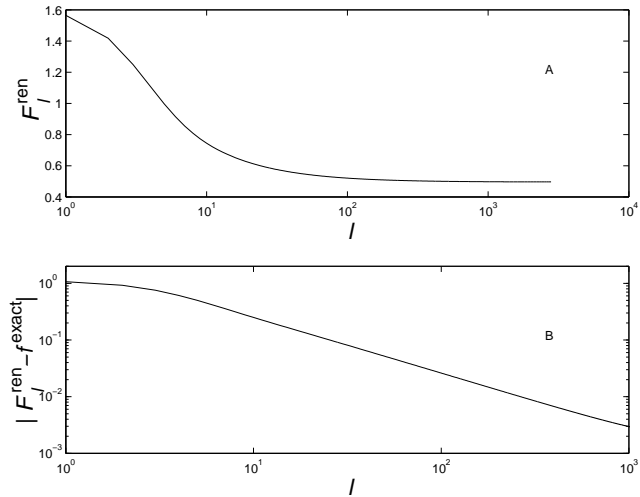}}
\caption{Top panel (A): The renormalized force ${^{\rm ren}}F^{l}_{r}$ as
a function of $l$. 
Bottom panel (B): $|F^{\rm ren}_l-f^{\rm exact}|$ as a function
of $l$. The charge is located at $r=2.1M$. The regularization procedure
is performed with $l'=2.8\times 10^{3}$.}
\end{figure}
 
\section{Summary}\label{discussion}
We presented a direct calculation of the self forces acting on two types
of static charges in Schwarzschild spacetime: a scalar charge and an
electric charge. In both cases the boundary conditions were chosen such
that the scalar field and the potential, correspondingly, would be regular
both at infinity and at the black hole's event horizon. Our method is
based on decomposition of the field and the force into modes. Each mode
satisfies an
ordinary differential equation which we solve exactly in terms of Legendre
Functions (in the scalar case) or derivatives of the Legendre functions
(in the electric case). We find the total bare forces by summing over all
modes
numerically. This total force typically diverges. 
We then regularize the divergent self force with two independent
procedures:
First, we model the point-like particle to be spatially extended, 
and then consider a sequence of such particles, letting the spatial
extention decrease. The divergent piece of the force is removed by a mass
renormalization procedure (i.e., it is used to redefine the mass of the
particle), and the remaining force approaches the self force in the limit
of vanishing extension.    
Second, we use Ori's mode-sum regularization prescription, and remove the
divergent piece of the force by studying the behavior of the bare force at
large values of the mode number, and subtracting the value of the bare
force at the limit of infinite mode numbers from all modes. Both
regularization procedures recovered the well known results for static
charges in the spacetime of a Schwarzschild black hole: a zero self force
in the scalar case, and a repelling radial self force in the electric
case. 

When one
compares the relative effectiveness of the two regularization
procedures, one finds that their effectivenesses are comparable.  
Specifically, for comparable values of $l$ up to which we sum  the series,
we find that for both regularization schemes we obtain 
similar deviations of the computed regularized forces from the exact
solutions, with roughly the same computation time. 

Evidently, more work is of need for both regularization prescriptions. In
particular, it is not presently understood how to apply MSRP for more
complicated cases, e.g.,  it is not presently clear whether there
are cases with non-vanishing parameters $c_{\mu}$, and whether the 
formalism can be extended to handle such cases (the radial component
$c_{r}$ was shown to be zero only for a scalar charge, although for  all
orbits in Schwarzschild). Also, it is not clear when
non-zero functions $d_{\mu}$ should be expected. A generalization of MSRP
to include also the gravitational case is also needed, a case for which
the inherent gauge problem should be solved. We are currently using MSRP
to study more complicated cases, in particular the self force acting on a
scalar charge in circular orbit around a Schwarzschild black hole 
\cite{burko-preparation}.   

\section*{Acknowledgments}
I have benefited from useful discussions with Jolien Creighton. I thank
Amos Ori for many stimulating discussion and for letting me use his
results before their publication. 
This research was supported by NSF grant AST-9731698 and NASA grant
NAG5-6840.

\begin
{appendix}

\section{Mode-sum regularization prescription (MSRP)}

In this Appendix we describe very briefly the main ideas behind 
Ori's method for regularizing the mode-sum (MSRP) 
\cite{ori-private,barack-ori} for a scalar charge in Schwarzschild. 
  
We emphasize that the work on this method is still in progress. 
However, for the case of a static scalar charge in Schwarzschild, the 
regularization scheme has been developed in full.

As was pointed out by Quinn and Wald \cite{quinn-wald97}, 
the physical self-force is the sum of two parts: (i) A local, 
Abraham-Lorentz-Dirac type term, and (ii) a ``tail'' term 
${^{\rm tail}}F_{\mu}$, associated with the tail part of the Green's 
function. The local term is trivial to calculate (and it anyway vanishes 
in the static case considered in this paper). We shall therefore consider 
here the tail term only. This term may be expressed as
\begin{equation}
{^{\rm tail}}F_{\mu}\equiv\lim_{\epsilon\to 0^{-}}{_{\epsilon}}F_{\mu},
\end{equation}
where ${_{\epsilon}}F_{\mu}$ denotes the contribution to the force 
(evaluated at $\tau =0$) from the part $\tau \leq \epsilon$ of 
the particle's world line. Decomposing this expression into $\ell$-modes, 
one finds
\begin{equation}
{^{\rm tail}}F_{\mu}=
\lim_{\epsilon\to 0^-}\sum_{\ell}{_{\epsilon}}f^{\ell}_{\mu}
=\lim_{\epsilon\to 0^{-}}\sum_{\ell}(
{^{\rm bare}}f^{\ell}_{\mu}-\delta{_{\epsilon}}f^{\ell}_{\mu}).
\label{a2}
\end{equation}
Here, ${_{\epsilon}}f^{\ell}_{\mu}$, $\delta{_{\epsilon}}f^{\ell}_{\mu}$, 
and ${^{\rm bare}}f^{\ell}_{\mu}$ denote the force from the
$\ell$-multipole of the field sourced by the interval $\tau \leq \epsilon$, 
the interval $\tau > \epsilon$, and the entire world line, respectively. 
The force 
${^{\rm bare}}f^{\ell}_{\mu}$ may be identified with the sum over
$m$ and $\omega$ 
of the contributions from all stationary Teukolsky modes $\ell,m,\omega$ 
for a given $\ell$ 
(recall that in calculating a stationary field's mode $\ell,m,\omega$ 
one takes 
the source term to be the {\it entire} world line). Since we are using 
the retarded Green's function, the part $\tau > 0$ does not contribute. 
However, the interval from $\epsilon$ to $0^+$ does contribute. 
Essentially, it is this part which is responsible to the instantaneous, 
divergent, piece of the Green's function, which should be removed 
from the expression for ${^{\rm tail}}F_{\mu}$.). 

A clarification is required here concerning the meaning of the last 
equality in Eq. (\ref{a2}): Let $r_0$ denote the value of $r$ at the
evaluation point. Then, ${_{\epsilon}}f^{\ell}_{\mu}$ is well-defined 
at $r=r_0$. The situation with ${^{\rm bare}}f^{\ell}_{\mu}$ and 
$\delta{_{\epsilon}}f^{\ell}_{\mu}$ is more involved, however.  
Each of these quantities has a well-defined value at  
the limit $r\to r_0^-$, and a well-defined value at the limit $r\to r_0^+$. 
Generically, for the $r$-component (and in some cases also for 
other components) these two one-sided values are not the same. Equation 
(\ref{a2}) should thus be viewed as an equation 
for either the limit $r\to r_0^-$ of the relevant quantities 
(i.e., ${^{\rm bare}}f^{\ell}_{\mu}$ and 
$\delta{_{\epsilon}}f^{\ell}_{\mu}$), or 
the limit $r\to r_0^+$ of these quantities. 
Obviously, this equation is also valid for the {\it averaged} force, i.e.,  
the average of these two one-sided values. In what follows we shall 
always consider the averaged force. Of course, the final result of 
the calculation, ${^{\rm tail}}F_{\mu}$ (which has a well-defined value 
at the evaluation point), is the same regardless of whether  
it is derived from its one-sided limit $r\to r_0^-$, 
or from $r\to r_0^+$, or from their average.

Next, we seek an $\epsilon$-independent function $h^{\ell}_{\mu}$, 
such that the series 
$\sum_{\ell}({^{\rm bare}}f^{\ell}_{\mu}-h^{\ell}_{\mu})$
converges.
Once such a function is found, then Eq. (\ref{a2}) becomes 
\begin{equation}
{^{\rm tail}}F_{\mu}=\sum_{\ell}({^{\rm bare}}f^{\ell}_{\mu}
-h^{\ell}_{\mu})
-\lim_{\epsilon\to 0^-}\sum_{\ell}
(\delta{_{\epsilon}}f^{\ell}_{\mu}-h^{\ell}_{\mu}).
\label{a3}
\end{equation}
In principle, $h^{\ell}_{\mu}$ can be 
found by investigating the asymptotic behavior of 
${^{\rm bare}}f^{\ell}_{\mu}$ as $\ell \to \infty$. It is also 
possible, however, to derive $h^{\ell}_{\mu}$ from the 
large-$\ell$ asymptotic behavior of 
$\delta{_{\epsilon}}f^{\ell}_{\mu}$ (the latter and 
${^{\rm bare}}f^{\ell}_{\mu}$ must have the same large-$\ell$ 
asymptotic behavior, because their difference yields a convergent 
sum over $\ell$). In addition to $h^{\ell}_{\mu}$, the investigation 
of $\delta{_{\epsilon}}f^{\ell}_{\mu}$ should also yield the parameter 
$d_{\mu} \equiv \lim_{\epsilon\to 0^-}\sum_{\ell}
(\delta{_{\epsilon}}f^{\ell}_{\mu}-h^{\ell}_{\mu})$, required for the 
calculation of ${^{\rm tail}}F_{\mu}$ in Eq. (\ref{a3}).

Since we only need the asymptotic behavior of 
$\delta{_{\epsilon}}f^{\ell}_{\mu}$ for arbitrarily small $|\epsilon|$, 
it is possible to analyze it using local 
analytic methods. In particular, we can apply a perturbation 
analysis to the $\ell$-mode field equation (in the time domain). 
That is, we express the $\ell$-mode 
effective potential $V^\ell(r)$ as a small perturbation 
$\delta V^\ell(r)$ over the value of $V^\ell(r)$ at the evaluation point, 
$V^\ell_0\equiv V^\ell(r=r_0)$. Expressing
$G^\ell[x^{\mu},x^{\mu}_{s}(\tau
)]$, the $\ell$-mode Green's function,  as a function of $\tau$ and
$z\equiv \tau\ell$, the perturbation analysis provides an expression for 
$G^\ell[x^{\mu},x^{\mu}_{s}(\tau )]$ as a power series in $\tau$ (with
$z$-dependent coefficients). Only terms 
up to order $\tau^2$ are required for the calculation of the 
self force (recall that eventually we take the limit 
$\epsilon \to 0$), and the perturbation analysis yields  
explicit expressions for the required three expansion coefficients 
of $G^\ell$ (as functions of $z$). Constructing
$\delta{_{\epsilon}}f^{\ell}_{\mu}$ 
from $G^\ell$ (essentially by integrating 
the latter's gradient from $\epsilon$ to $\tau=0$), it can be shown 
that the large-$\ell$ asymptotic behavior of 
$\delta{_{\epsilon}}f^{\ell}_{\mu}$ takes the form  
$\delta{_{\epsilon}}f^{\ell}_{\mu}=  
a_{\mu} \ell +b_{\mu} +c_{\mu}\ell ^{-1}+O(\ell^{-2})$, 
in which the parameters $a_{\mu},b_{\mu},c_{\mu}$ 
are independent of $\ell$ and 
$\epsilon$ (though they depend on the orbit and evaluation point). 
(It can also be shown that there is no logarithmic divergence of
$h^{\ell}_{\mu}$.) 
The regularization function $h^{\ell}$ thus takes the form 
$h^{\ell}_{\mu}=a_{\mu} \ell +b_{\mu} +c_{\mu}\ell^{-1}$, and 
the tail part of the self force is given by
\begin{equation}
{^{\rm tail}}F_{\mu} 
=\sum_{\ell}\left({^{\rm bare}}f^{\ell}_{\mu}
-a_{\mu}\ell-b_{\mu}-c_{\mu}\ell^{-1}\right)-d_{\mu}.
\end{equation}

In the case of a static scalar particle in Schwarzschild, 
one can show that $a_{\mu}=c_{\mu}=d_{\mu}=0$ \cite{ori-private}. 
($a_{\mu}$ and $c_{\mu}$ are likely to vanish for all orbits in
Schwarzschild, but so far this has been shown explicitly for the 
radial component only.) 
The self-force for a static particle then takes the simple form
\begin{equation}
{^{\rm tail}}F_{\mu}
=\sum_{\ell}({^{\rm bare}}f^{\ell}_{\mu}-b_{\mu}).
\end{equation}
Namely, in this simple case the regularization procedure
is reduced to subtracting ${^{\rm bare}}f^{\ell\to\infty}_{\mu}$, 
the large-$\ell$ limit 
of the $\ell$ multipole of the bare force, from each multipole $\ell$ 
(note that since $a_{\mu}=0$, 
$b_{\mu}\equiv{^{\rm bare}}f^{\ell\to\infty}_{\mu}$).
For the particular case of a static scalar charge in Schwarzschild, Ori
\cite{ori-private} also obtained analytically the value of this
large-$\ell$ 
limit of the force: 
$b_{\mu}=-[q^2/(2r^2)](1-M/r)/(1-2M/r)\delta_{\mu}^{r}$.  

This regularization prescription takes a trivial form in the cases of
static scalar or electric charges in Minkowski spacetime. In these cases
it is easy to verify that all the $\ell$ modes of the bare force are equal 
(i.e., independent of $\ell$), 
specifically ${^{\rm bare }}f_r^{\ell}=-q^2/(2r^2)={\rm const}$ (this can
be obtained easily directly from a decomposition of the field). 
When one sums over all modes, the
bare force of course diverges. However, subtracting this constant term
from each mode yields a new series, where all modes are zero, such that
the total force vanishes, which is the well-known result in Minkowski
spacetime. We note that MSRP turns out to be
effective also for the cases of scalar or electric charges in circular
orbits in Minkowski spacetime \cite{paperI}.

We emphasize that whereas the parameters $a_{\mu}$, $b_{\mu}$, and
$c_{\mu}$
can be found from the behavior of ${^{\rm bare}}f^{\ell}_{\mu}$ at 
large values of the mode number $\ell$, the parameter $d_{\mu}$ can only
be calculated according to its definition. In the simple case of a 
scalar charge in circular orbit around Schwarzschild, which includes as
a special case a static scalar charge, this calculation is not  
hard to do, and the exact value of $d_{\mu}$ was found (for this
case $d_{\mu}=0$). However, it might be the case that for more complicated
cases $d_{\mu}$ is more difficult to find. Then, one can still regularize
the force, but the result would not be free from the ambiguity which  
results from the ignorance of the exact value of $d_{\mu}$.

MSRP involves integration over the entire world
line of the orbiting object. In that respect, it is especially suitable  
for periodic, or near periodic, orbits. For a-periodic orbits, such as the
final plunge of the object into the black hole, one can perhaps use a
different approach, where one integrates 
only over the past world line, excluding the position of the particle
itself, and thus avoids the singular contribution. 
Of course, the closer to
the particle one integrates, the more modes one would need to sum over in
order to obtain convergence. In fact, the number of modes is inversely
proportional to the proper-time difference from the event up to which one
integrates and the position of the particle.
It has been recently shown by Wiseman \cite{wiseman-unpublished} that in
the far-field limit (i.e., to leading order in the ratio of the black hole
mass to the radius of the orbit), the contribution of the near
neighborhood of the past world line is negligible, such that one may need 
to sum only over a relatively small number of modes. In stronger fields,
this approach can perhaps be combined with a normal neighborhood expansion
\cite{wiseman-unpublished,anderson-flanagan} to obtain the self force.

\section{Numerical evaluation of the Legendre functions}
All the series we need to evaluate very accurately involve the Legendre
functions of the first and second kinds, and their derivatives. The degree 
of the functions is very high. For example, in the numerical summations
reported here we evaluated the series up to $l=4.5\times 10^{4}$. This
requires
us to use a very accurate algorithm for the calculation of the Legendre
functions. In fact, we find that the Legendre functions of the first kind
can be very accurately computed by using the recursion relations, such as 
\begin{equation}
P_{l+1}(x)=\frac{1}{l+1}\left[(2l+1)xP_{l}(x)-lP_{l-1}(x)\right]$$
\end{equation}
and
\begin{equation}
P'_{l}(x)=\frac{l}{x^2-1}\left[ xp_{l}(x)-P_{l-1}(x)\right] .$$
\end{equation}
Although similar relations hold also for the Legendre function of the
second kind, they are not practical for the following reason. The
functions $Q_{l}(x)$ approach zero very fast for fixed $x>1$ when the
degree gets very large. The subtraction which is inherent to the recursive
expression becomes numerically inaccurate very rapidly. The functions
$Q_{l}$ can also be considered as the sum of two series, one being a
polynomial, and the other being a polynomial multiplied by a common
logarithmic factor. Each of the polynomials satisfy the same recursive
formula as the Legendre functions, but with different initial terms for
$l=0$ and $l=1$. Each of the
two series grow very fast with $l$, but their difference gets very small.
Therefore, this method would also not be very accurate numerically. A way
to avoid these difficulties is to use the integral representation of the
functions $Q_{l}(x)$. This is given by 
\begin{equation}
Q_{l}(x)=\frac{1}{2^{l+1}}\int_{-1}^{1}\,dt\frac{(1-t^2)^l}{(x-t)^{l+1}}.
\end{equation}
The integrand does not have any pathologies in the entire interval of
integration, and also the boundaries are regular. We perform this integral
using Romberg integration, which proves to be very efficient and accurate 
\cite{numerical-recipes}. The derivatives of the functions $Q_{l}(x)$ can
still be computed by the relation given above for $P'_{l}(x)$. Another
improvement on the numerical evaluation of both $P_{l}(x)$, $Q_{l}(x)$ and
their derivatives is the following. In all the expressions we have, we
need only compute the {\it product} of two Legendre functions or their
derivatives, one factor involving $P_{l}$ (or its derivative), and the
other involving $Q_{l}$ (or its derivative). Because we are not interested
in the value of the Legendre functions themselves, but only in such
products, we can disregard the factor of $2^{-(l+1)}$ in the integral
representation of the $Q_{l}(x)$. This would mean that each of the
functions we compute is too large by a factor of $2^{l+1}$. If we then
compute, instead of the functions $P_{l}(x)$, a new function, which is
smaller than $P_{l}(x)$ by the same factor, the product of the two new
functions would be unchanged. This can be done also for the derivatives of
the Legendre functions. It is advantageous to do this, because for a given
floating point arithmetic this procedure increases the maximal value of
$l$ for which accurate computations can be performed by an order of
magnitude.

\end{appendix}

\end{document}